\documentstyle[epsfig]{mn}{}
\begin{document}
\def\mpc{h^{-1} {\rm{Mpc}}} 
\def\up{h^{-3} {\rm{Mpc^3}}} 
\def\uk{h {\rm{Mpc^{-1}}}}
\def\kms {\rm{km~s^{-1}}} 
\def\apj {ApJ} 
\def\aj {AJ} 
\def\mnras {MNRAS} 
\title{Systems of Galaxies in the SDSS: the fundamental
plane} 
\author[Eugenia D\'{\i}az and Hern\'an Muriel] {Eugenia
D\'{\i}az and Hern\'an Muriel\\ Grupo de Investigaciones en
Astronom\'{\i}a Te\'orica y Experimental, IATE, Observatorio
Astron\'omico, Laprida 854, C\'ordoba, Argentina\\ Consejo de
Investigaciones Cient\'{\i}ficas y T\'ecnicas de la Rep\'ublica
Argentina (CONICET)} 
\date{\today} \maketitle 
\begin{abstract} 
We analyse a subsample of the galaxy groups 
obtained by Merch\'an \& Zandivarez (2005) from the SDSS DR3 to study
the fundamental plane and the mass to light ratio of galaxy groups.
We find a fundamental plane given by $L_R \propto R^{1.3}
\sigma^{0.7}$. We do not find differences when different dynamical
sates or redshift ranges are analysed. We find that the mass to light
ratio increases with group mass as $M/L_R \propto M^{0.36}$.
\end{abstract} 
\section{Introduction} 
The study of the early type galaxies has
allowed the discovery of a plane in the 3-D space of intrinsic
properties of galaxies. This plane is known as the fundamental plane
(FP) and is expressed as the relation between luminosity, size and
intrinsic kinetic energy (Dressler et al., 1987; Djorgovski \&
Davis, 1987; Guzm\'an et al., 1993). From the analysis of the FP,
information about physical properties, formation and evolution of
systems can be obtained. Moreover, the FP has been extensively used
as a distance indicator playing an important role in the
determination of the Hubble constant ($H_0$). 

The FP concept has also
been extended to other systems such as galaxy clusters. Schaeffer et
al. (1993), Adami et al. (1998), Fujita \& Takahara (1999) and
Fritsch \& Buchert (1999) have confirmed the existence of a
fundamental plane for these large systems. Another topic to be
considered when the FP is analysed is the dynamical state of the
sample. Fritsch \& Buchert (1999) claim that clusters with less
substructures (more relaxed) are the strongest tracers of the FP and
suggests that the dispersion around the FP is the result of systems
of galaxies with a lower degree of relaxation. Beyond these
preliminary results, all these authors agree that a larger sample is
necessary to have significant statistical weight. 

At present, the
largest redshift survey of galaxies is the Sloan Digital Sky Survey
(SDSS) DR3. Recently, Merch\'an \& Zandivarez (2005) have
identified groups of galaxies in this survey, providing the largest
sample of groups. Using a sub-sample of this group catalogue, the
present work studies the fundamental plane of galaxy groups and their
mass to light ratio. The outline of this paper is as follows: in
section 2, we describe the data sample; in section 3, we briefly
describe the set of parameters used to define the fundamental plane,
while the fit itself is presented in section 4. The mass to light
ratio analysis is detailed in section 5. We summarise our results and
conclusions in section 6.
\section{The data sample} 
The present work is based on a subsample of the groups
identified in the SDSS DR3 by Merch\'an \& Zandivarez (2005). 
Due to the nature of the present
work, a very reliable and homogeneous sample of groups is required.
Therefore, we only select those groups with at least 10 members.
Since the parameters that define the FP can be sensitive to the
selection of the groups centre, we implemented the iterative method
described by D\'{\i}az et al. (2005), which reduces the contamination
by substructure. The final sample (hereafter MZDM sample) consists 
of $495$ groups. The
median redshift, 3-D velocity dispersion and number of members are $0.077$, 
$642 \ km \ s^{-1}$ and $14$, respectively. 
The distribution of velocity dispersions
shown in the left panel of Figure \ref{fig0} indicates that our sub-sample includes
both low and high mass systems of galaxies.
\section{The set of parameters} 
\subsection{Optical luminosity} 
The luminosity of a group of galaxies 
identified within a magnitude-limited galaxy sample needs to be corrected for
incompleteness effects. In order to correctly compute the luminosity
of each group identified in the MZDM sample we use the method described by
Moore et al.(1993) . According to these authors, the group optical
luminosity is defined by the following expression: 
\begin{equation}
L_R=L_g+L_{corr} \end{equation} where \begin{equation}
L_g=\sum_{i=1}^{Ngal}L_{i} \end{equation} with
$L_{i}=10^{M_i-M_\odot}$, and \begin{equation} L_{corr}=N_{gal}
\frac{\int^{L_{lim}}_0 L_R \Phi_R(L) dL}{\int^{\infty}_{L_{lim}}
\Phi_R(L) dL} 
\end{equation} where
$L_{lim}=10^{0.4(M_\odot-M_{lim})}$.  $\Phi_R(L)$ is the luminosity
function of galaxies in groups. Throughout this work we use
luminosities in the R-band. 

\begin{figure}
\epsfxsize=0.5
\textwidth 
\hspace*{-0.5cm}
\centerline{\epsffile{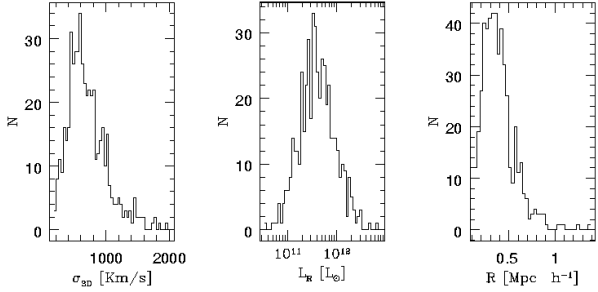}} 
\caption{From left to right:
distribution of 3-D velocity dispersions, R-band luminosities and
radius. 
\label{fig0}} 
\end{figure}

The absolute magnitudes $M_i$ are
calculated using the $k+e$ corrections as a function of redshift,
following a method similar to that described by Norberg et al.
(2002). This method uses the Bruzual \& Charlot (1993) stellar
population synthesis code. The luminosity functions of galaxies in
groups are estimated following the procedure described by
Mart\'{\i}nez et al. (2002). Using the complete sample of galaxies in
groups identified by Merch\'an \& Zandivarez (2005), we found the
following Schechter parameters: $\alpha=-1.00 \pm 0.03$ and
$M*=-20.57 \pm 0.04$. The adopted absolute solar magnitude is
$M_\odot=4.62$ (Blanton et al. 2003). The middle panel of 
Figure \ref{fig0} shows the
distribution of our group luminosities which extends from $3.41
\times 10^{10} L_\odot$ to $6.94 \times 10^{12} L_\odot$. We adopt an 
upper limit to the measurement error of $15 \%$ in the luminosities 
as recommended by Adami et al.(1998). 
\subsection{Velocity dispersions and radius} 
The velocity dispersion 
of each group is calculated using the standard technique described by Beers et al.
(1990). We apply the biweight estimator for groups with richness
$N_{tot} \ge 15$ and the gapper estimator for poorer groups. The
median 3-D velocity dispersion for the complete sample of groups is $
(642 \pm 190)\ km \ s^{-1}$. The error in the 3-D velocity dispersion
is around of $30 \%$ as stated by Beers et al.(1990). 

The group
characteristic radii is calculated as suggested by Eke et al.(2004a).
These authors compute the projected group size using the rms
projected physical separation of the galaxies respect to the group
centre: 
\begin{equation} R=\sqrt{\frac{\sum_{j=1}^{Ngal} \
d^2_{jc}}{Ngal}} 
\end{equation} 
where $d_{jc}$ is the projected
distance between the centre position and the $j^{th}$ galaxy and Ngal
is the number of group members. The right panel of Figure \ref{fig0} shows the
distribution of group radius. The median radii of the sample is
$(0.36 \pm 0.10) \ Mpc \ h^{-1}$. 
The error in the radius $R$ was estimated using Monte Carlo realisations of mock groups with a given density profile. The procedure takes into account the number of members used to compute the radius and includes uncertainties in the centre position. Considering possible differences between mock and real groups, we adopt a 20\% as a conservative upper limit for the error in the group’s radius. 
\section{The fundamental plane} 
\begin{figure} 
\epsfxsize=0.5
\textwidth
\hspace*{-0.5cm} 
\centerline{\epsffile{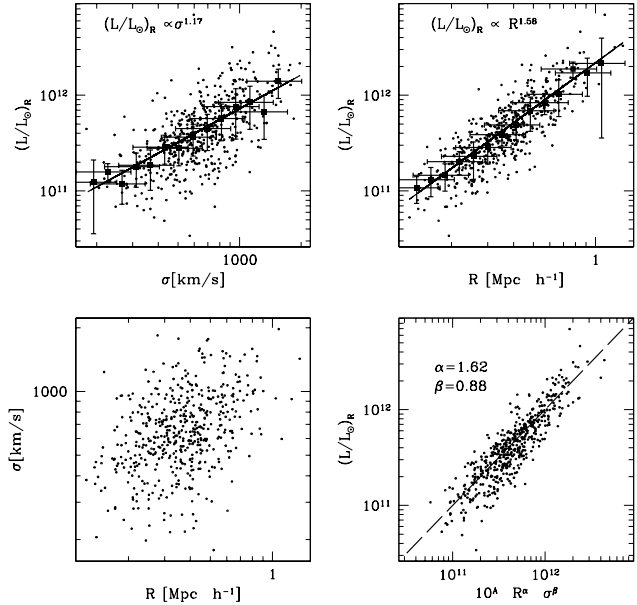}} 
\caption{Left upper
panel: $L_R-\sigma$ relation. Solid line is the best fit relation.
Right upper panel: $L_R-R$ relation. Solid line is the best fit.
Filled squares in both panels correspond to the median luminosities
per bin of velocity dispersion (left upper) or radius(right upper).
Left lower panel: $\sigma$-R relation. Right lower panel:
$L_R-\sigma-R$ relation. The abscissa is $10^A R^\alpha
\sigma^\beta$, where A, $\alpha$ and $\beta$ are the best fit
parameters (see text). 
\label{fig1}} 
\end{figure} 

A simple way to
start the study of the fundamental plane is by analysing its
different projections. Figure \ref{fig1} shows the $L_R-\sigma$,
$L_R-R$ and $R-\sigma$ projections of the FP. As can be appreciated
in the left and right upper panels, both, the $L_R-\sigma$ and the
$L_R-R$ show a clear correlation in the sense that groups that have
large radii or high velocity dispersions tend to be more luminous
than those that are smaller or dynamically colder. We fit the
$L_R-\sigma$ relation using a method that minimises the sum of the
squared weighted orthogonal distances to an analytical curve (or
surface). Throughout this work, we perform the fitting procedures
using the routines of ODRPACK (Boggs et al., 1992), which takes into 
account errors in all the coordinates involved. The errors assigned to each 
coordinate are: $\epsilon _L/L=0.15$, $\epsilon _R/R=0.2$, 
and $\epsilon_\sigma/\sigma=0.3$ . The best fit is
shown in the left upper panel of Figure \ref{fig1} (solid line).
Filled squares are the median luminosities per bin of velocity
dispersion, the error associated with the median is the
semi-interquartile range. The best fitting relation is:
\begin{eqnarray} (L/L_\odot)_R=10^{b_1} \ \sigma ^{a_1}
\label{l-sigma} 
\end{eqnarray} 
with $a_1= 1.17 \pm 0.09$ and
$b_1=8.35 \pm 0.25$. The right upper panel shows the $L_R-R$
relation, the best fit is: 
\begin{eqnarray} (L/L_\odot)_R=10^{b_2} \
R^{a_2} \label{l-R} 
\end{eqnarray} 
where $a_2=1.58 \pm 0.06$ and
$b_2=12.34 \pm 0.03$. 

Left lower panel of Figure \ref{fig1} shows the
projection of the FP in the $R-\sigma$ plane. Larger groups tend to
have higher velocity dispersions; however, the correlation is
marginal. Due to the poor correlation, the fitting routine does not
produce an acceptable relation between radius and velocity
dispersion. 

Finally, the right lower panel shows the $L_R-\sigma-R$
relation.
The fit to the data corresponds to the plane equation
\begin{eqnarray} (L/L_\odot)_R=10^{A} \ R^{\alpha} \ \sigma^{\beta}
\label{FP} 
\end{eqnarray} 
The best-fitting parameters given by the ODRPACK subroutines are:
$\alpha=1.32 \pm 0.06$, $\beta=0.70\pm 0.05$ and $A=10.3\pm 0.2$.

Even though a good correlation is found, one of the key questions is
the origin of the observed dispersion, which could be a consequence
of the contribution of groups with different characteristics. Several
authors have found that clusters lie in a plane in the 3-D space of
$L-\sigma-R$. Nevertheless, they still discuss how the fundamental
plane must be defined. Should all the groups lie in the same plane?
Or, is the fundamental plane only well defined for groups with some
particular physical properties? The assumption of virial state
implies that clusters have a constant mass to light ratio, which
suggests that groups should lie in a plane defined by $L \propto R
\sigma^2$. Nowadays, we know that not all the clusters are
virialized, and that the dynamical equilibrium is less common in
groups. A more realistic determination of the dynamical state of
groups is thus necessary. The size of our group sample gives us a
unique opportunity to test whether group dynamical state is one of
the factors responsible for the observed dispersion. 
\begin{figure}
\epsfxsize=0.3
\textwidth 
\hspace*{-0.5cm}
\centerline{\epsffile{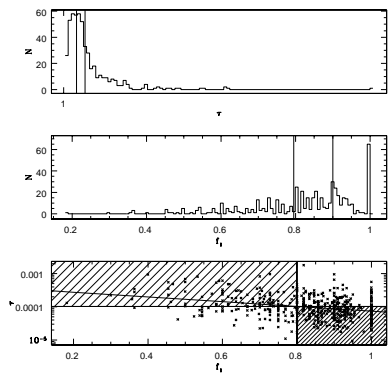}} 
\caption{Upper panel: $\tau$
distribution. Vertical lines show the $\tau_1$ and $\tau_2$ values.
Middle panel: $f_1$ distribution. Vertical lines are the $frac_1$ and
$frac_2$ values. Lower panel: correlation $\tau$ vs $f_1$. Wide
hatched region corresponds to the less evolved groups. Narrow hatched
region corresponds to the more evolved groups. 
\label{fig2}}
\end{figure} 

The dynamical state of a group can be studied in
different ways. Taking into account the available information, we
apply two complementary parameters: a dimensionless crossing time,
$\tau$, and the early type fraction in groups, $f_1$. 

(i) the
dimensionless crossing time, used by Hickson et al (1992), reflects
the dynamical evolution since it is proportional to the inverse of
the number of times that a galaxy could have traversed the group from
its formation to the present time. $\tau$ is defined by:

\begin{eqnarray} 
\tau=H_0 \ t_c= \frac{400}{\pi} \
\frac{\Delta}{\sigma} \label{tau} 
\end{eqnarray} 
where $\Delta$ is
the mean projected galaxy separation in a group, and $\sigma$ is the
3-D velocity dispersion. 

(ii) If the morphology of galaxies in groups
and clusters are the result of environmental processes that
subsequently transform galaxies between different morphological
classes, early type galaxies should be more numerous in evolved
clusters than in young less evolved systems.The fraction of early
type galaxies per group is computed after splitting the galaxy sample
into 3 spectral types, following D\'{\i}az et al. (2005). The
fraction $f_1$ is: $f_1=N_1/N$, where $N$ and $N_1$ group total
number of members and the number of early type galaxies,
respectively. $f_1$ should reflect the degree of relaxation of a
system. 

Neither $\tau$ nor $f_1$ are strongly correlated with the
redshift nor with the group mass, which is calculated following Eke
et al. (2004a): 
\begin{eqnarray} M=5 \frac{\sigma^2 \ R}{G}
\label{energy} 
\end{eqnarray} 

We study the dependence of the
fundamental plane on these dynamical parameters. First at all, we
define subsamples according to their corresponding $\tau$ values: 
(1) more evolved: $\tau \le \tau_1= 7.6 \times 10^{-5}$, 
(2) intermediate evolution: $\tau_1 < \tau \le \tau_2=1.26 \times 10^{-4}$, 
and (3) less evolved: $\tau > \tau_2$. Upper panel of Figure \ref{fig2}
shows the $\tau$ distribution. Vertical lines are the $\tau_1$ and
$\tau_2$ values. We fit a plane (eq. \ref{FP}) for each subsample. We
find no differences between these planes and the defined by the whole
sample. We measure the orthogonal scatter around the FP. This
orthogonal scatter quantify the aloofness from the FP. Neither of the
3 subsamples shows differences in the scatters. 

We split the group
sample into the following subsamples, according to their fraction of
early type galaxies : (1) less evolved $f_1 \le frac_1=0.795$,
(2)intermediate evolution $frac_1< f_1 \le frac_2=0.9$, and (3)
more evolved $f_1>frac_2$. The quoted values of $frac_1$ and
$frac_2$ where selected in order to have subsamples of similar size.
The middle panel in Figure \ref{fig2} shows the $f_1$ distribution,
and the $frac_1$ and $frac_2$ values. Applying the same analysis used
for $\tau$, we find no differences in the fitted FPs as in the
orthogonal dispersions for the three different subsamples. 

Finally,
we seek for a correlation between $\tau$ and $f_1$. It is shown in
the lower panel of Figure \ref{fig2}. It can be seen that the
relation is not injective. We perform a linear fit in logarithmic
axes which is shown as solid line in the Figure. Then, we combine
both parameters to pick up two subsamples, corresponding to the more
(narrow hatched region) and less (wide hatched region) evolved
groups: (1) $\tau \le \tau_*=1.03 \times 10^{-4}$ and $f_1 > f_*=0.8$, and (2)
$\tau > \tau_*$ and $f_1 \le f_*$. Again, we compute the plane and
the orthogonal scatters around the FP for each subsample. The results
are the same found before. Both subsamples have the same behaviour.

From the analysis performed in this section, we conclude that,
using the parameters $\tau$ and $f_1$ to study the group dynamical
state, the fundamental plane does not show signs of evolution. 

We also study the dependence of the fundamental plane on the group
redshifts. We define 2 subsamples corresponding to the lowest and the
highest redshift ranges. The resulting planes have no differences
with the FP defined by the whole sample, and the orthogonal scatters
around the FP are very similar for both subsamples. However, this
result is not conclusive since our sample spans only a small redshift
range ($z \le 0.2$), wherein only minor dynamical evolution is
expected. 

\begin{figure}
\epsfxsize=0.4\textwidth \hspace*{-0.5cm}
\centerline{\epsffile{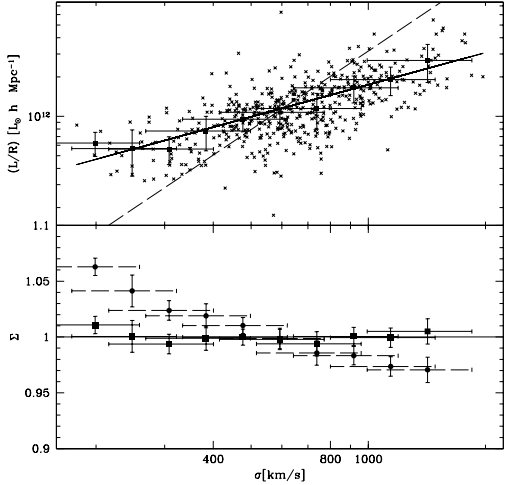}} \caption{Upper panel: $(L/R)-\sigma$
relation. Filled squares are the median values per bin of $\sigma$.
Solid line is the best fit to the median data points. Dashed line is
the relation expected from the virial equilibrium assumption. Lower
panel: Ratios between median values and the linear relations. Filled
squares corresponds to best fit, and filled circles are computed with
the virial prediction. 
\label{fig3}} 
\end{figure}

Finally, in order to show that the fundamental plane
expected from the virial assumption is rejected by the MZDM sample,
the upper panel of Figure \ref{fig3} shows the $(L/R)-\sigma$
relation. Solid line is the best fit to the data (filled squares:
$(L/R)_{median}$ per bin of $\sigma$), and dashed line is the
relation expected when assuming virial state. Lower panel of this
Figure shows the ratios between the median values and the linear
relations (best fit: filled squares - virial relation: filled
circles). It can be seen that the virial relation is not a good
description to the observational data. 

\begin{table*}
\begin{center}
\caption{Fundamental plane and $L-M$ best-fitting parameters for different prescriptions of size parameter and mass.}
\begin {tabular}{ccccccccc}
\hline 
\hline 
Prescription &  \multicolumn{3}{c}{$L=10^A \ R^\alpha \ \sigma ^\beta$} & & \multicolumn{1}{c}{FP orthogonal scatter} & & \multicolumn{2}{c}{$L=10^b \ M ^a$} \\
\cline{2-4} \cline{6-6} \cline{8-9}
 & $\alpha$ & $\beta$ &  A  & & $\Sigma$ & & a & b  \\
\hline 
 Eke et al.(2004a) &  $1.32\pm0.06$  &  $0.70\pm0.05$  &  $10.3\pm0.2$ & & $0.08$& &$0.64\pm0.03$ & $2.6\pm0.4$  \\
\hline 
 Virial & $1.85\pm0.08$  & $0.73\pm0.06$  & $9.7\pm0.2$ & & $0.10$ & & $0.64\pm0.04$ &$2.7\pm0.6$  \\
\hline 
\end{tabular}
\end{center}
\end{table*}
\section{Mass to light ratio} 
\begin{figure}
\epsfxsize=0.4
\textwidth 
\hspace*{-0.5cm}
\centerline{\epsffile{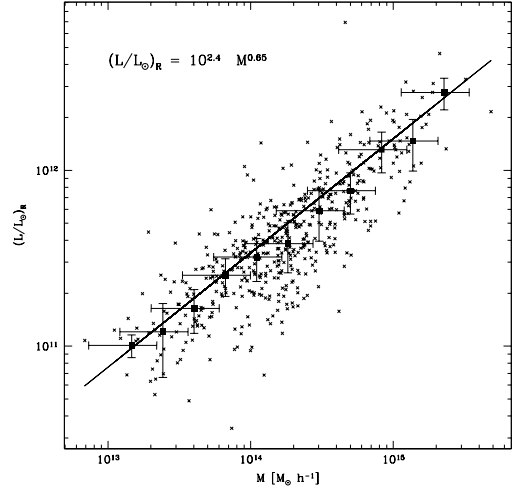}} 
\caption{$L_R-M$ relation. Filled
squares are the median luminosity per mass bin. Solid line is the
best fit. 
\label{fig4}} 
\end{figure} 

The fact that the FP we measure
is different from the one expected assuming virial equilibrium ($L
\propto R \sigma^2$) means that the mass to light ratio must vary.
Girardi et al.(2000) calculate the $L_B-M$ relation and find that the
luminosity has a tendency to increase slower than the mass ($L_B
\propto M^{0.75}$). These authors also suggest that this result is
independent of the photometric band, which was confirmed by Popesso et
al. (2004). Figure \ref{fig4} shows the $L_R-M$ scatter plot (points)
corresponding to the complete sample of groups. Filled squares are
the median luminosity per bin of mass, errors in the median
luminosities are computed as the semi-interquartile range, and the
mass errors are computed by error propagation ($\sim 60 \%$). Solid
line corresponds to the best fit to $L_R=10^b M^a$, with $a=0.64 \pm
0.03$ and $b=2.6 \pm 0.4$. This result is in agreement (within 2
$\sigma_a$) with the results obtained by Girardi et al. (2000), and it
is also comparable (within $\sigma_a$) with the results of Popesso et
al. (2004). It should be noted that $L$ varies almost linear with
$\sigma$($\beta \sim 1$) (quadratic in the virial case), then the
$M/L$ ratio must increase with $\sigma$, it means with $M$. Several
authors have stated that is not correct to search for the best
fitting relation of $M-L$ ratio versus $M$ or $L$ (Eke et al., 2004b, Popesso
et al., 2004, Girardi et al. 2002), then it is more suitable to
infer the relations from the $L \ vs \ M$ directly. Therefore, our
previous result implies $M/L \propto M^{0.36 \pm 0.06}$, it means
that the mass to light ratio of galaxy groups is not constant, 
$M/L$ varies up to a factor of $\sim 6$ from low to high mass
groups. 
The group sample analysed in our work presents a steeper slope of the $M/L$ vs 
$M$ relation, in comparison with previous works on groups and clusters of 
galaxies ($0.25\pm0.1$
Girardi et al., 2000, Adami et al. 1998; $0.2\pm 0.08$ Popesso et al.
2004), but they are in good agreement within 1 $\sigma$-level. The median
mass to light ratio of our sample is $(M/L)_{med}=(418\pm194)
M_\odot/L_\odot$. 

In order to check the stability of our results against a different choice 
of the group size, we repeated our analysis using the group standard 
virial radius and the virial mass provided by Merch\'an \& 
Zandivarez 2005.
The median virial radius of the sample under study is $0.96\pm0.20
\ Mpc \ h^{-1}$. The relation between the virial radius and the
radius used in this work is linear ($R_{vir}=(1.72\pm0.02) \ R +(0.29\pm0.01)$)
with a small dispersion, but it has a non zero intercept. Then, the fundamental
plane fitted using the virial radius slightly differs from the fit
derived in the previous section.

The $L-M$ relation does not depend on the definition of the radius or mass. 
Comparing the orthogonal scatter produced by the two different selection 
of the size parameters, we find that the characteristic radius proposed 
by Eke et al. (2004a) produces the smaller scatter in both, 
the fundamental plane and the $L-M$ fits, and it also produces 
smaller errors in the fitted parameters. 
Table 1 shows the results corresponding to the two
size parameters.

\section{Conclusions}
In this work we study whether the more numerous galaxy systems, the
galaxy groups, lie in the so-called "fundamental plane",
defined by their physical properties. We analyse 
a subsample of the Merch\'an \& Zandivarez (2005) catalogue of groups 
(MZDM sample).
The use of this large and
homogeneous sample allows us to obtain results that are statistically
reliable. We find that these groups define a plane given by $L_R
\propto R^{1.3} \ \sigma ^ {0.7}$ which is different from the plane
that is expected if one assumes virial equilibrium. We also analyse
the aloofness from the plane as a function of the dynamical state of
groups and their redshifts.We find that none subsample has a
tendency to lie farther or closer from the FP. 

We also find that the
mass to light ratio increases with group mass as $(M/L_R) \propto
M^{0.36}$. Over the mass range of our sample (two orders of
magnitude), the $M/L_R$ ratio increases a factor of $\sim 6$ from low
to high mass systems. 
\section{Acknowledgements}
We thanks to the anonymous referee for helpful suggestions that
improved this work. We thanks to Darren Reed for reading the
manuscript and useful comments. We also thank to SDSS Team for having
made available the actual data sets of the sample. This work has been
partially supported by Consejo de Investigaciones Cient\'{\i}ficas y
T\'ecnicas de la Rep\'ublica Argentina (CONICET), the Agencia
Nacional de Promoci\'on Cient\'{\i}fica, the Secretar\'{\i}a de
Ciencia y T\'ecnica de la Universidad Nacional de C\'ordoba (SeCyT).

\end{document}